\documentclass[openacc]{rsproca_new}

\usepackage[british]{babel}

\usepackage{anyfontsize}
\usepackage{rjwmath}
\usepackage{mathrsfs}  
\DeclareMathOperator{\rect}{rect}
\DeclareMathOperator{\sinc}{sinc}
\DeclareMathOperator*{\argmax}{argmax}

\usepackage{braket}

\newcommand{\kdom}{k_{\text{dom}}}

\titlehead{Research}

\begin{document}


\title{Evolving wrinkles: time-dependent buckling of an elastic sheet on a liquid substrate}

\author{
Daniel J.\ Netherwood$^{1}$, Ben S.\ Humphries$^{2}$, Connor Robbins$^{3}$, Doireann O'Kiely$^{4}$}

\address{$^{1}$ School of Computer and Mathematical Sciences, University of Adelaide, Adelaide, SA 5005, Australia\\
$^{2}$ School of Engineering, Mathematics, and Physics, University of East Anglia, Norwich, NR4 7TJ, UK\\
$^{3}$ School of Environmental Sciences, University of East Anglia, Norwich, NR4 7TJ, UK \\
$^{4}$ Mathematics Applications Consortium for Science and Industry (MACSI), Department of Mathematics and Statistics, University of Limerick, Ireland\\
}

\subject{Applied Mathematics, Mechanics}

\keywords{Wrinkling, Pattern formation, Fluid-structure interaction}

\corres{Daniel J. Netherwood\\
\email{daniel.netherwood@adelaide.edu.au}}

\begin{abstract}
We model the formation and evolution of wrinkles in a floating elastic sheet under uniaxial compression.  This is a canonical setup in the study of wrinkling, and whilst its static equilibrium configuration is well characterised, its dynamics are not.  In this work, we focus on modelling the transition from early, inertia-dominated wrinkle growth to late-time gravity-moderated equilibrium.
For an initial configuration in which the sheet is flat, an initial disturbance will first grow at the shortest available wavelengths, because this requires the least kinetic energy, but will subsequently transition to a longer preferred wavelength that minimises potential energy.  We observe that the evolving wave pattern must be a spectrum, as opposed to a fundamental wrinkle mode whose wavelength evolves in time. Our results demonstrate that changes in the dominant wrinkle wavelength are coupled to a decay in the compressive force, which is to be expected from equilibrium theory.
\absbreak
As part of this study, we found that the system must have a dissipation mechanism in order to evolve into an equilibrium steady state, and that the dominant wrinkle wavelength and compressive force are independent of this mechanism.  Instead it is found that dissipative effects influence the shape of the wavepacket, with the spectrum narrowing as the dissipation increases.
\end{abstract}

\rsbreak 

\section{Introduction}
Wrinkles occur in a plethora of real-world scenarios, both naturally and through human engineering. In the natural world, examples include gyrification of the human brain during fetal development \cite{greiner2021physical,balbi2020mechanics,karzbrun2018human}, growth-induced buckling of epithelial cell layers \cite{youn2024tissue,edwards2007biomechanical,rozman2021morphologies} and the formation and expansion of biofilms \cite{geisel2022role,trejo2013elasticity,fei2020nonuniform}. In engineering, wrinkles can be used to redirect light~\cite{Harrison2004,Lee2014}, modify surface properties~\cite{Chung2009}, or facilitate flexibility in electronic components~\cite{RogersHuang2009}; these wrinkles can be formed by, for example, compressing a sheet that is attached to a substrate.  In all cases, wrinkles form because they are an energy-efficient mechanism for a material to accommodate excess length or area, and the wrinkle formation occurs at an emergent wavelength that balances the competing forces within a given system.  In particular, the bending stiffness of the sheet itself resists high-curvature (short-wavelength, small-amplitude) configurations, while substrate stiffness, tension and curvature along wrinkles, and confining walls, all resist large-displacement (long-wavelength, large-amplitude) configurations.  More complex patterns can also be generated through bi-axial confinement; for example,  the shrinking associated with curing or drying~\cite{Vandeparre2010} and the stresses induced by forcing a naturally curved surface to be flat~\cite{Tobasco2022} can both lead to visually striking multi-directional wrinkles. 

The competing effects that determine wrinkle patterns offer routes for controlling pattern selection.  For example, decreasing the stiffness of the sheet and increasing substrate stiffness will both act to decrease the characteristic size of the wrinkles, while modifying applied forces gives opportunities for tuning the wrinkle pattern in a given sheet-substrate system~\cite{LandauLifschitz,CerdaMahadevan, Paulsen2016}. 
 The general rule for wrinkle wavelength selection is neatly summarised in the literature~\cite{LandauLifschitz,CerdaMahadevan, Paulsen2016} as
 \begin{equation}\label{eq:ClassicalWavelength}
\lambda^* = 2 \pi \left( \frac{D^*}{K^*} \right)^{1/4},
 \end{equation}
 in equilibrium, where $D^*$ is the bending stiffness of the material and $K^*$ is a resisting stiffness, which is equal to density $\times$ gravity if the substrate is a hydrostatic fluid, and can be adapted for more complicated scenarios.  
 Allowing for dynamics introduces an additional layer of questions:  how does a wrinkled configuration evolve to its equilibrium state, and what are the factors that control this evolution?  Experimental investigations have pointed to viscous dissipation, substrate inertia and sheet inertia as additional factors that can affect how wrinkles evolve towards their final state \cite{Vandeparre2010,box2019dynamics, box2020dynamic}. %
 
Processes like polymer curing motivate studies where wrinkling is dynamic due to a viscous substrate~\cite{Vandeparre2010}.  For example, compression of a thin elastic sheet on a lubricated layer can be described mathematically and yields wrinkles whose wavelength grows slowly in time $t^*$ according to~\cite{kodio2017lubricated} 
 \begin{equation}
     \lambda^* \propto \left(\frac{t^*}{\log t^*}\right)^{1/6}.
 \end{equation}
 Practically, a tri-layer system composed of a pre-stretched elastic base, a viscous layer, and a thin elastic sheet gives a route for probing viscosity-mediated wrinkling experimentally~\cite{chatterjee2015}.  When the pre-stretch in the base elastic layer is released, liquid in the viscous layer is pulled inwards, which in turn drags the thin layer on top inwards, inducing wrinkles whose development must displace the viscous fluid further.  Wrinkle localisation and ridge formation can also occur~\cite{guan2023ridge,guan2023flat}.

 While the scenarios mentioned in the paragraph above are viscosity-dominated, in other scenarios the dynamics of wrinkle formation over a fluid are affected by inertia.  Indentation of a floating sheet is one example of a system where this can be studied.  Indentation of the sheet centre induces hoop stresses that in turn drive the formation of spoke-like wrinkles;  for quasistatic indentation the wavelength of these wrinkles decreases as the indentation increases~\cite{Vella2015PRL, Vella2018PRE, Ripp2020}.  At high indentation speeds, the inertia of the fluid delays the formation of large-wavelength wrinkles, and so at early times the wrinkle wavelength grows from vanishingly small, with a trend close to~\cite{box2019dynamics,okiely2020impact}
 \begin{equation}
 \lambda^* \propto \left(t^*\right)^{2/5}.
 \end{equation}
 However, modelling the dynamics beyond very early times is complicated by a variety of factors associated with the sheet shape and size as well as geometric and material nonlinearities.   
 
 If the dynamics of the substrate are removed, then dynamic effects can come from the sheet itself, e.g. from the inertia of the wrinkling elastic material.  One way of studying this experimentally is to indent or contract an elastic sheet in freefall~\cite{VVV2007,VVV2009, duchemin2014impact, box2020dynamic, kodio2020dynamic}.  This opens an additional angle for wrinkles to behave dynamically.

In this paper we consider wrinkling of an elastic beam floating on a liquid substrate, accounting for both inertia and viscosity, with the goal of resolving the wrinkle evolution from initiation to equilibrium.  We avoid the problems encountered previously for indentation of a circular sheet by moving to a more theoretically convenient scenario:  a beam (or infinitely wide sheet) compressed uniaxially by prescribed displacement of its ends. This paper is organised as follows. In Section~\ref{sec:Problem_description}, we outline the mathematical model and present the various parameter regimes applicable to this study. In Section~\ref{sec:analysis}, we nondimensionalise the problem before exploiting the parameter regime in order to derive a relatively simple mathematical description of a beam subject to hydrodynamic, hydrostatic and viscous forces. A solution of this reduced problem is then sought by performing a model decomposition by Fourier transform in space, which yields a system of evolution equations for the amplitudes of different wrinkling modes with different wrinkle wavelengths.  This facilitates a quantification of the length, kinetic energy and potential energy associated with each mode.  The evolution equations are coupled via a length constraint, forming a differential algebraic system which requires a careful numerical approach, as outlined in Section~\ref{sec:numericalmethod}.  We present numerical solutions and interpret our findings in Section~\ref{sec:results} and draw conclusions in Section~\ref{sec:conclusions}.

\section{Mathematical model}\label{sec:Problem_description}
\subsection{Problem description}
\begin{figure}
\centering
{
\fontsize{22pt}{24pt} 
\resizebox{100mm}{60mm}{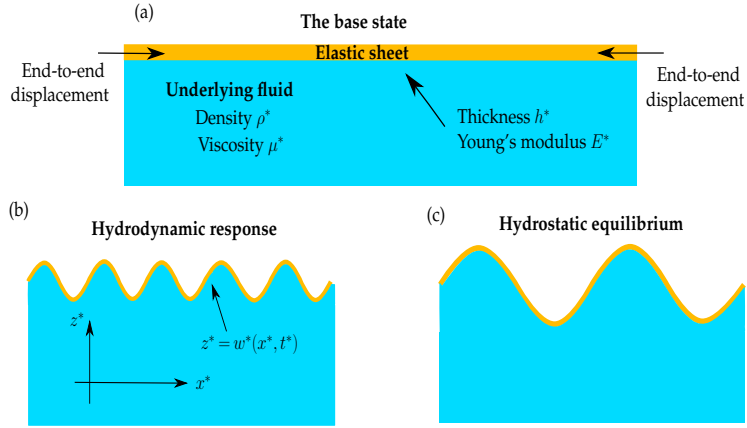}
}
\caption{(a) The base-state configuration of an elastic beam, which at time $t^* = 0$ is subject to a prescribed end-to-end displacement of size $\Delta^*$, which buckles the beam out-of-plane, displacing the underlying viscous fluid leading to (b) a highly-wrinkled beam configuration, induced by viscous and inertial restoring forces. As time evolves, the wrinkles coarsen, eventually leading to (c) the late-time steady state for which hydrostatic and spring restoring forces overcome visco-inertial forces and combine to balance bending effects.}\label{fig:Setup}
\end{figure}

We consider a long, thin and linearly elastic beam of length $2L^*$ and Young's modulus $E^*$, which in its base-state configuration (see figure \ref{fig:Setup}a) is flat, and lies on the surface of a hydrostatic fluid of density $\rho^*$ and dynamic viscosity $\mu^*$ occupying the semi-infinite domain $(x^*,z^*) \in (-L^*,L^*) \times (-\infty, 0)$. In its base-state configuration, the beam is located at $z^* = 0$ and occupies the domain $x^* \in (-L^*,L^*)$. We consider the case in which both endpoints of the beam are instantaneously displaced inwards by an amount $\Delta^*$. In response, the beam buckles with out-of-plane displacement $w^*$, which in turn excites an unsteady viscous flow in the underlying fluid. The precise asymptotic regime under consideration as well as the dimensional equations governing both the fluid flow and the beam's displacements are presented in the following two sections. 

\subsection{The asymptotic regime}\label{sec:AsymptoticRegimeSection}
We consider an asymptotic regime in which the beam's characteristic displacement amplitude
$W^*$ is small when compared with the characteristic axial lengthscale $\ell^*$, which will later be chosen as the emergent late-time steady-state wrinkle wavelength. Hence, we define the wrinkle aspect ratio
\begin{equation}
    \epsilon = \frac{W^*}{\ell^*} \ll 1.
\end{equation}

Our next task is to quantify the relative importance of linear inertial effects, nonlinear inertial effects and viscous effects.  In this work, we assume that linear inertial effects are important, but also that the effects associated with nonlinear inertia may be neglected, and that viscous effects are weak but non-negligible.  The importance of nonlinear versus linear inertial effects is quantified by the parameter $\epsilon \ll 1$.  For viscous effects, the standard Reynolds number $Re = \rho^* V^* \ell^*/\mu^*$ (with $V^*$ defined to be the characteristic magnitude of the fluid velocity) quantifies the importance of \emph{nonlinear} inertial effects relative to viscous effects, so it is helpful to introduce a timescale $T^* = W^*/V^*$ and define a modified Reynolds number $Re/\epsilon$ which quantifies \emph{linear} inertial effects relative to viscous effects.  That is, the overall importance of viscous effects is measured by
\begin{equation}
\mu = \frac{\mu^* V^* /\ell^{*2}}{\rho^* V^*/T^*} = \frac{W^*}{\ell^*} \frac{\mu^*}{\rho^* \ell^* V^*} = \frac{\epsilon}{Re}.
\end{equation}
Our regime choice then corresponds to the distinguished limit in which $\mu \to 0$ as $\epsilon,Re \to 0$, or alternatively
\begin{equation}\label{eq:regime}
\epsilon \ll \mu \ll 1.
\end{equation}

\subsection{Fluid mechanics}
The fluid velocity $\bm{v}^*$ and hydrodynamic pressure, $p^*$ are governed by the Navier--Stokes equations in the fluid bulk, with a dynamic boundary condition at the interface that equates stresses in the fluid with those in the beam.  For the regime $\epsilon \ll \mu \ll 1$ specified in Section~\ref{sec:AsymptoticRegimeSection}, we follow the work of Dias \emph{et al.}~\cite{dias2008theory} to simplify the fluid-mechanical governing equations by resolving the fluid velocity in terms of scalar and vector potentials $\phi^*$ and $\bm{\psi}^* = \psi^* \bm{j}$ via a Helmholtz decomposition
\begin{equation}\label{eq:HelmholtzDecompMain}
    \bm{v}^* = \nabla^* \phi^* + \nabla^* \times (\psi^* \bm{j}) = \left(\pdiff{\phi^*}{x^*}-\pdiff{\psi^*}{z^*}\right)\bm{i} + \left(\pdiff{\phi^*}{z^*}+\pdiff{\psi^*}{x^*}\right)\bm{k},
\end{equation}
where $(\bm{i}, \bm{j}, \bm{k})$ are the usual Cartesian unit vectors. Using this decomposition, it is shown in Appendix \ref{sec:AppendixDias} that mass and momentum conservation are described by
\begin{equation}\label{eq:DimensionalLaplacesAndDiffusionEquation}
    {\nabla^*}^2 \phi^* = 0, \qquad \mu^* {\nabla^*}^2 \psi^* = \rho^*\pdiff{\psi^* }{t^*},
\end{equation}
respectively, while stress continuity at the fluid--beam interface is described by
\begin{equation}\label{eq:DimensionalViscoInertialDynamicCond}
     \rho^*\pdiff{\phi^*}{t^*} + \rho^* g^* w^* + 2 \mu^* \left(\frac{\partial^2 \phi^*}{\partial {z^*}^2}+\frac{\partial^2 \psi^*}{\partial z^* \partial x^*}\right) = -p_{\mathrm{ext}}^* \qquad \text{at} \qquad z^* = w^*,
\end{equation}
for the normal stress component, where $p^*_{\mathrm{ext}}$ is the pressure exerted on the beam by the fluid, and 
\begin{equation}\label{eq:DimensionalTangentialStressCond}
    2\left(\frac{\partial^2 \phi^*}{\partial z^* \partial x^*}-\frac{\partial^2 \psi^*}{\partial {z^*}^2}\right) +\frac{\rho^*}{\mu^*}\pdiff{\psi^*}{t^*} = 0 \qquad \text{at} \qquad z^* = w^*
\end{equation}
for the tangential stress component.
Finally, it is shown that the continuity of normal velocities at the fluid-beam interface gives rise to the kinematic condition
\begin{equation}\label{eq:DimensionalKinematicMain}
    \pdiff{\phi^*}{z^*}+ \pdiff{\psi^*}{x^*}  =\frac{\partial w^*}{\partial t^*}+\left(\frac{\partial \phi^*}{\partial x^*}-\frac{\partial \psi^*}{\partial z^*}\right)\frac{\partial w^*}{\partial x^*} \qquad \text{at} \qquad z^* = w^*.
\end{equation}
Our model also requires prescription of boundary and initial conditions. We impose an infinite-depth condition 
\begin{equation}
    \frac{\partial \phi^*}{\partial z^*} \to 0 \qquad \text{as} \qquad z^* \to -\infty
\end{equation}
on the fluid velocity.  We assume that the beam length is sufficiently large that finite-length boundary effects may be ignored, rendering the $x^*$-axis effectively infinite. Formally, the required condition is then that the fluid velocity remains bounded in the far field. We postpone discussion of the initial conditions to \S\ref{sec:InitialConds}.

\subsection{Beam mechanics}
Turning our attention to the beam mechanics, under the small-strain assumption, the out-of-plane displacement and axial tension in the beam are governed by the Föppl-von Kármán equations
\begin{align}
     \frac{\partial{F}^*}{\partial{x}^*} &=0, \label{eq:Dimensional_FplVonK1} \\
    D^*\frac{\partial^4 w^*}{\partial {x^*}^4} - F^*\frac{\partial^2 w^*}{\partial {x^*}^2} &= p_{\mathrm{ext}}^*, \label{eq:Dimensional_FplVonK2}
\end{align}
where $F^*=F^*(t^*)$ is the thickness-integrated time-dependent axial stress.  It is noted that here we have assumed that the beam is in stress equilibrium, which requires beam inertia to be negligible. This is valid provided that the mass per unit thickness of the beam is small when compared with the mass per unit wavelength of the fluid~\cite{OKielyVella2025}.

Eliminating the external pressure between (\ref{eq:DimensionalViscoInertialDynamicCond}) and (\ref{eq:Dimensional_FplVonK2}) yields
\begin{equation}\label{eq:FSIBeameq}
    \rho^* \frac{\partial \phi^*}{\partial t^*} + D^*\frac{\partial^4 w^*}{\partial {x^*}^4} - F^*(t^*)\frac{\partial^2 w^*}{\partial {x^*}^2} + \rho^* g^* w^* + 2 \mu^* \left(\frac{\partial^2 \phi^*}{\partial {z^*}^2}+\frac{\partial^2 \psi^*}{\partial z^* \partial x^*}\right) = 0
\end{equation}
at $z^* = w^*$. Equation (\ref{eq:FSIBeameq}) is to be appended to the governing equations (\ref{eq:DimensionalLaplacesAndDiffusionEquation}) for the fluid flow in the bulk as a dynamic boundary condition that captures the fluid-structure interaction.

The final governing equation, which may be interpreted as a closure condition for the time-dependent eigenvalue, $F^*$, is given by 
\begin{equation}\label{eq:DimensionalLengthConstraint}
    \frac{1}{2}\int_{-L^*}^{L^*} \left(\pdiff{w^*}{x^*}\right)^2 \D{x}^* = 2\Delta^*.
\end{equation}
Equation (\ref{eq:DimensionalLengthConstraint}) is a geometrically nonlinear length constraint on the beam that normalises the out-of-plane displacement, $w^*$, by ensuring that it absorbs all of the excess length available for wrinkling induced by the end-to-end displacement.
\section{Mathematical analysis}\label{sec:analysis}
\subsection{Scalings and nondimensionalisation}\label{sec:ScalingAnalysis}
In this section we discuss variable scalings and present the nondimensionalisation of equations (\ref{eq:DimensionalLaplacesAndDiffusionEquation})--(\ref{eq:DimensionalLengthConstraint}). We begin by using (\ref{eq:DimensionalLaplacesAndDiffusionEquation}) to motivate the introduction of a common horizontal and vertical lengthscale $(x^*\sim z^*)$, denoted $\ell^*$. Balancing the restoring force due to bending with hydrostatic/buoyancy restoring forces in (\ref{eq:FSIBeameq}), $\ell^*$ is chosen to be
\begin{equation}\label{eq:FFTlengthscale}
     \ell^* = \left(\frac{D^*}{\rho^* g^*}\right)^{1/4}.
\end{equation}
On this lengthscale, the terms in (\ref{eq:FSIBeameq}) associated with axial bending and tension-curvature balance when the scale for the tension is chosen to be 
\begin{equation}\label{eq:FFTTensionScale}
    F^* \sim \sqrt{D^* \rho^* g^*}.
\end{equation}
The scales (\ref{eq:FFTlengthscale})--(\ref{eq:FFTTensionScale}), which form the basis of our choice of nondimensionalisation of equations (\ref{eq:DimensionalLaplacesAndDiffusionEquation})--(\ref{eq:DimensionalLengthConstraint}), are classical results in tension-field theory, see e.g.~\cite{CerdaMahadevan, TaffetaniVella2017, OKielyVella2025}. Thus we scale
\begin{align}
    & w^* = W^* w, \qquad (x^*,z^*) = \ell^*(x,z), \qquad (\phi^*,\psi^*) = \frac{\ell^* W^*}{T^*}\left(\phi,\mu\psi\right), \nonumber \\ & \qquad p^* = \rho^* g^* W^* p, \qquad t^* = T^*t, \qquad F^* = \sqrt{D^* \rho^* g^*}F,\label{eq:NonDimensionalisation}
\end{align}
where the scale for the out-of-plane displacement $W^* = \ell^* \sqrt{\Delta^*/L^*}$ has been chosen on examination of (\ref{eq:DimensionalLengthConstraint}), and where the explicit timescale $T^* = \sqrt{\ell^*/g^*}$ has been obtained by balancing linear fluid inertia with hydrostatic forces in (\ref{eq:FSIBeameq}). Finally, we comment that our choice of scaling for $\psi^*$ in (\ref{eq:NonDimensionalisation}) arises from analysing the tangential stress condition (\ref{eq:DimensionalTangentialStressCond}), which suggests that it is natural to take $\psi^* \sim \epsilon Re^{-1}\phi^* = \mu \phi^*$.
\subsection{Model reduction and dimensionless system of governing equations}
We now substitute the nondimensionalisation (\ref{eq:NonDimensionalisation}) into the system (\ref{eq:DimensionalLaplacesAndDiffusionEquation})--(\ref{eq:DimensionalLengthConstraint}), and simplify the resulting dimensionless system by retaining terms that are leading order with respect to the small parameter $\epsilon$. We preserve weak dissipative effects by retaining contributions that are first order with respect to $\mu \ll 1$. The dimensionless governing equations to be solved in the fluid domain are
\begin{equation}\label{eq:PotentialsGovEqDimensionless}
    \nabla^2 \phi = 0, \qquad \mu \nabla^2 \psi = \pdiff{\psi}{t}, \qquad \text{on} \qquad (x,z) \in (-L,L) \times (-\infty, 0),
\end{equation}
where $L = L^*/\ell^*$ and we anticipate taking the limit $L \to \infty$ below. Equations (\ref{eq:PotentialsGovEqDimensionless}) are to be solved subject to the far-field boundary condition 
\begin{equation}
    \pdiff{\phi}{z} \to 0 \qquad \text{as} \qquad z\to -\infty,
\end{equation}
together with the requirement that the fluid velocity remains bounded at the beam's endpoints $|x|=L$. At the linearised fluid-beam interface, we have the leading-order kinematic condition
\begin{equation}\label{eq:DimensionlessKinematicCond}
     \pdiff{w}{t} = \pdiff{\phi}{z} + \mu \pdiff{\psi}{x} \qquad \text{at} \qquad z = 0. 
\end{equation}
The normal and tangential stress conditions are
\begin{equation}\label{eq:DimensionlessForceBalance}
     \pdiff{\phi}{t} + \pdiff[4]{w}{x} - F(t) \pdiff[2]{w}{x} + w + 2 \mu \pdiff[2]{\phi}{z} = 0 \qquad \text{at} \qquad z = 0,
\end{equation}
and 
\begin{equation}\label{eq:DimensionlessTangentialStressCond}
    2\frac{\partial^2 \phi}{\partial z \partial x} +\pdiff{\psi}{t}-2\mu \pdiff[2]{\psi}{z} = 0\qquad \text{at} \qquad z = 0,
\end{equation}
respectively.  We interpret the normal stress condition~\eqref{eq:DimensionlessForceBalance} as a F\"{o}ppl-von K\'{a}rm\'{a}n equation with hydrodynamic pressure.  The amplitude of $w$ is set by the geometric constraint 
\begin{equation}\label{eq:DimensionlessLengthConstraint}
    \frac{1}{2}\int_{-L}^{L}\left(\pdiff{w}{x}\right)^2 \D{x} = 2L.
\end{equation}
It is noted that in steady state there is no fluid velocity and~\eqref{eq:DimensionlessForceBalance} reduces to 
\begin{equation}\label{eq:staticw}
    \pdiff[4]{w}{x} - F \pdiff[2]{w}{x} + w = 0,
\end{equation}
which is the F\"{o}ppl-von K\'{a}rm\'{a}n equation for a beam with unit bending, on a substrate with unit restoring stiffness, subjected to an axial tension $F$.  It is well known that~\eqref{eq:staticw} has solutions of the form $\mathrm{e}^{ikx}$ with $k\in \mathbb{R}$ for compressive forces $F\leq -2$, and that the solution $k=1$ with $F=-2$ minimises the sum of bending and substrate energies subject to the constraint~\eqref{eq:DimensionlessLengthConstraint}~\cite{CerdaMahadevan,Paulsen2016}.

\subsection{Modal decomposition}\label{sec:fourier}
By linearity in the governing equations (\ref{eq:PotentialsGovEqDimensionless})--(\ref{eq:DimensionlessTangentialStressCond}), it is natural to seek a solution by Fourier transform in $x$ so that a separable solution takes the form:
\begin{align}
    \phi(x,z,t) &= \sqrt{\frac{L}{\pi}} \int_{-\infty}^\infty\frac{1}{|k|}f(k,t) \mathrm{e}^{ikx} \mathrm{e}^{|k|z}\D{k}, \label{eq:FT_phi} \\ 
    \psi(x,z,t) &= \sqrt{\frac{L}{\pi}} \int_{-\infty}^\infty\frac{1}{ik} g(k,t)\mathrm{e}^{ikx}\mathrm{e}^{|m|z}\D{k},\label{eq:FT_psi} \\
    w(x,t) &= \sqrt{\frac{L}{\pi}} \int_{-\infty}^\infty a(k,t)\mathrm{e}^{ikx}\D{k},\label{eq:FT_w}
\end{align}
where $k\in \mathbb{R}$ is the wrinkle wavenumber and $m = m(k)$. The inclusion of the prefactors within the integrands of (\ref{eq:FT_phi})--(\ref{eq:FT_psi}) are motivated by the kinematic condition (\ref{eq:DimensionlessKinematicCond}), and the geometric prefactors in each of (\ref{eq:FT_phi})--(\ref{eq:FT_w}) have been included for convenience when substituting the transforms into the length constraint (\ref{eq:DimensionlessLengthConstraint}). We note at this stage that whilst the beam's dimensionless length $L$ is finite, it shall here and henceforth be taken as sufficiently large that finite-length effects at the beams endpoints may be ignored, effectively rendering the beam infinite. 

Substituting the transforms~\eqref{eq:FT_phi}--\eqref{eq:FT_w} into the mass conservation equation~\eqref{eq:PotentialsGovEqDimensionless} and the kinematic condition~\eqref{eq:DimensionlessKinematicCond}, we find that $a$, $f$ and $g$ are related by:
\begin{equation}\label{eq:fk_gk_ak_relationships}
    \pdiff{g}{t} = \mu\left(|m|^2-k^2\right)g(k,t), \qquad \pdiff{a}{t} = f(k,t)+\mu g(k,t).
\end{equation}
Similarly, the F\"{o}ppl-von K\'{a}rm\'{a}n equation (\ref{eq:DimensionlessForceBalance}) gives 
\begin{equation}\label{eq:akfull}
    \frac{1}{|k|}\left(\pdiff[2]{a}{t}-\mu \pdiff{g}{t}\right) +\left(k^4+Fk^2+1\right)a + 2\mu |k|\left( \pdiff{a}{t} - \mu g \right) = 0,
\end{equation}
where we have used the result~(\ref{eq:fk_gk_ak_relationships}) to eliminate $f$. Finally, noting that $\mu \, {\partial g}/{\partial t} = O(\mu^2)$ by (\ref{eq:fk_gk_ak_relationships}) and eliminating all terms at this order, the governing equation~\eqref{eq:akfull} for $a$ simplifies to
\begin{equation}\label{eq:ak_equation}
    \frac{1}{|k|}\pdiff[2]{a}{t} +  2\mu |k| \pdiff{a}{t} +  \left(k^4+Fk^2+1\right)a = 0.  
\end{equation}
Equation (\ref{eq:ak_equation}) is a second-order partial differential equation governing the the amplitude, $a(k,t)$, of the beam's out-of-plane displacements as a function of the wavenumber and time, as well as the axial tension, $F$ as a function of time. For brevity we refer to it as FvK below.
The closure condition enabling the determination of $F$ at each moment in time is obtained by substituting (\ref{eq:FT_w}) into the length constraint (\ref{eq:DimensionlessLengthConstraint}), and is found to be
\begin{equation}\label{eq:a_k_integralconstraint}
    \int_{-\infty}^\infty \frac{1}{2} k^2 |a|^2 \D{k} = 1 ,
\end{equation}
where $|a|^2 = a \overline{a}$ with $\overline{a}$ denoting the complex conjugate of $a$. Equations~\eqref{eq:ak_equation},~\eqref{eq:a_k_integralconstraint} together with suitable initial conditions are investigated numerically below.
\color{black}

\subsection{Energetic considerations and steady states}\label{sec:EnergeticConsiderations}
Before equipping the system \eqref{eq:ak_equation}--\eqref{eq:a_k_integralconstraint} with initial conditions and solving numerically, we first investigate the system's energetics to motivate the expected outcomes.  
In the absence of beam inertia, the only source of kinetic energy comes from the underlying fluid. In contrast, the system's potential energy has contributions from both the bending moment in the beam and the weight of the underlying fluid.  The total dimensionless energy can be decomposed as $E =  E_{\mathrm{kin}}+ E_{\mathrm{pot}}$, where the components $E_{\mathrm{kin}}$ and $E_{\mathrm{pot}}$ are written in terms of the spectrum~\eqref{eq:FT_w} as 
\begin{equation}\label{eq:system_energy_integral}
   E_{\mathrm{kin}}(t) = \int^{\infty}_{-\infty} \mathcal{E}_{\mathrm{kin}} \D{k}
   \qquad
   \text{and}
   \qquad
   E_{\mathrm{pot}}(t) = \int^{\infty}_{-\infty} \mathcal{E}_{\mathrm{pot}} \D{k},
\end{equation}
respectively, and 
\begin{equation}\label{eq:energy_density_expressions}
    \mathcal{E}_{\mathrm{kin}} (k,t)= \frac{1}{2|k|} \bigg|\pdiff{a}{t}\bigg|^2, \qquad  \mathcal{E}_{\mathrm{pot}} (k,t) =\frac{1}{2} \left( k^4  +1 \right)   |a|^2,
\end{equation}
are the respective kinetic and potential energy densities, which, along with their global counterparts, have been scaled with a factor $L^2/\pi$ for convenience.  These expressions can be derived from physical quantities or inferred from \eqref{eq:ak_equation}--\eqref{eq:a_k_integralconstraint} (see Appendix \ref{sec:AppendixEnergy} for details).  

Differentiating the total energy $E$ in time and making use of the governing equation \eqref{eq:ak_equation}, its complex conjugate, and the time-derivative of \eqref{eq:a_k_integralconstraint}, yields
\begin{equation}\label{eq:decreasingenergy}
    \diff{E}{t} = - 2\mu  \int^{\infty}_{-\infty}  |k| \bigg|\pdiff{a}{t}\bigg|^2 \D{k} \leq 0.
\end{equation}
Comparing~\eqref{eq:system_energy_integral} and~\eqref{eq:decreasingenergy}, we see that the total energy is positive but monotonically decreasing. If $\mu=0$ the total energy will remain at its initial value, if $\mu \neq 0$ then, by application of the monotone convergence theorem, it will decrease to some other positive constant. In the latter case (dissipative, $\mu \neq 0$), as $\mathrm{d}E/\mathrm{d}t$ vanishes so must the
the integrand on the right-hand side of~\eqref{eq:decreasingenergy}; this requires that $\partial a / \partial t \to 0$ and so the system eventually reaches a steady state. 

From the evolution equation~\eqref{eq:ak_equation}, the only possible steady states are those where $a$ has isolated peaks at the roots of $k^4 + Fk^2 + 1 = 0$, and such real roots exist if $F \leq -2$. 
From the static literature~\cite{CerdaMahadevan,Paulsen2016}, minimising the potential energy density with respect to $k$, subject to the length constraint~\eqref{eq:a_k_integralconstraint}, yields
\begin{equation}\label{eq:static}
k^2=1, \quad \text{with} \quad F=-2,
\end{equation}
as expected.  
It is natural to wonder whether a system initiated close to a different steady state (i.e. with $F < -2$ and a peak at $k^2 \neq 1$) might stay close to that steady state.  From a physical standpoint, any perturbation that excites modes that don't satisfy $k^4 + Fk^2 + 1 = 0$ will induce velocities and hence dissipation, driving it towards~\eqref{eq:static} provided $\mu \neq 0$.  More generally, from a mathematical viewpoint, we can ask if any steady state with $F < -2$ is a local minimiser of total energy.  To answer this question we study the energetic stability of the system by calling on variational calculus. Following classical theory, one formulates the appropriate energy functional (with constraints), before computing first variations to obtain the governing Euler--Lagrange equations. Second variations are then computed in order to assess the stability of equilibria \cite{gelfand_fomin,lanczos, pignataro_rizzi_luongo_1991}. 

We introduce a generalised energy functional
\begin{equation}
\hat{E}[a,b] = \frac{1}{2} \int^{\infty}_{-\infty} \left[ \frac{1}{|k|} |b|^2 + \left( k^4  +1 \right)   |a|^2 + F k^2 |a|^2 \right]\ \mathrm{d}k,
\end{equation}
where $b = \partial a/\partial t$ and the final term can be interpreted as imposing the length constraint or quantifying work done on the beam ends.  For a perturbation $[\delta a, \delta b]$ the first variation of the energy is  
\begin{equation}
\delta \hat{E} = \frac{1}{2} \int^{\infty}_{-\infty} \left[ \frac{1}{|k|} (b \delta \overline{b} + \overline{b} \delta b) + \left( k^4  + Fk^2 + 1 \right)  \left(a \delta \overline{a} + \overline{a} \delta a \right) \right]\ \mathrm{d}k,
\end{equation}
and the second variation is
\begin{equation}
\delta^2 \hat{E} = \frac{1}{2} \int^{\infty}_{-\infty} \left[ \frac{1}{|k|} |\delta b|^2 + \left( k^4 + Fk^2 + 1 \right)   |\delta a|^2 \right]\ \mathrm{d}k.
\end{equation} 
For any permitted steady state the first variation is zero and the sign of the second variation gives the stability.  A stable steady state is then one for which $\delta^2 \hat{E} \geq 0$ for any perturbation, and is only satisfied if $k^4 + Fk^2 + 1 \geq 0$, i.e. there is only one stable steady state, the one at $F = -2$.

Finally, it is noted that in the absence of dissipation ($\mu = 0$), the total energy is conserved and the $k^2=1$, $F=-2$ equilibrium outlined above cannot typically be reached.  Nevertheless, an increase in kinetic energy from its initial value will drive a decrease in potential energy, which corresponds to transferring length to modes closer to $k=1$.  

\section{Numerical analysis}\label{sec:numericalmethod}
\subsection{Numerical method}\label{sec:gekko}
Re-introducing $b(k,t)={\partial a}/{\partial t}$, the problem~\eqref{eq:ak_equation}--\eqref{eq:a_k_integralconstraint} can be restated as finding a solution set $a,b,F$ to the system
\begin{align}
    \pdiff{a}{t} &= b \label{eq:DAE_system_1}, \\
    \pdiff{b}{t} &= -2\mu k^2 b -  |k| \left(k^4+Fk^2+1\right)a, \label{eq:DAE_system_2}\\
    0 &= 1 - \int_{-\infty}^\infty \frac{1}{2}  k^2 |a|^2 \D{k} \label{eq:DAE_system_3} ,
\end{align}
given a set of initial conditions $ a(k,0),\, b(k,0),\, F(0) $.  
Owing to the integral constraint (\ref{eq:DAE_system_3}), which will become an algebraic constraint upon numerical discretisation, equations (\ref{eq:DAE_system_1})--(\ref{eq:DAE_system_3}) form what is canonically referred to as a system of differential-algebraic equations (DAEs) \cite{gear1971simultaneous, Campbell:2008, ascher1998computer}.
DAEs are typically classified by a so-called `differentiation index', defined to be the minimum number of times the algebraic constraint (\ref{eq:DAE_system_3}) is to be differentiated before it can be recast as an explicit first-order system \cite{campbell1995index}. 

Systems of DAEs with a single algebraic constraint with a differentiation index of $n > 1$ must, for consistency, satisfy $n-1$ additional constraints obtained by differentiating the algebraic constraint $n-1$ times \cite{ascher1998computer}.
 The index of the system \eqref{eq:DAE_system_1}--\eqref{eq:DAE_system_3} is three and hence we obtain two additional constraints upon computing the first and second derivatives of (\ref{eq:DAE_system_3}), namely
\begin{equation}\label{eq:firstderivative}
    \int_{-\infty}^\infty k^2 \left( a \overline{b} + \overline{a\vphantom{b}} b \right) \D{k} = 0,
\end{equation}
and
\begin{equation}\label{eq:secondderivative}
    \int_{-\infty}^\infty k^2 \left[ |b|^2 - |k|\left(k^4 + Fk^2 +1\right)|a|^2 - \mu k^2 \left( a \overline{b} + \overline{a\vphantom{b}} b\right)   \right] \D{k} = 0.
\end{equation}
The final constraint~\eqref{eq:secondderivative} can be rearranged to yield an expression for $F$, given as
\begin{equation}\label{eq:F_formula}
    F =  \frac{\int_{-\infty}^\infty k^2 \left[ |b|^2 - |k|\left(k^4 +1\right)|a|^2 - \mu k^2 \left(a \overline{b} + \overline{a \vphantom{b}} b  \right) \right] \D{k}}{\int_{-\infty}^\infty k^4 |k| |a|^2 \D{k}}.
\end{equation}
Differentiating (\ref{eq:F_formula}) once in time and combining the result with (\ref{eq:DAE_system_1})--(\ref{eq:DAE_system_2}) yields an explicit first-order system for $(a,b,F)$, hence verifying that the differentiation index of (\ref{eq:DAE_system_1})--(\ref{eq:DAE_system_3}) is three.

We seek a numerical solution of the system (\ref{eq:DAE_system_1})--(\ref{eq:DAE_system_3}) by employing the method of lines. Firstly, it is noted that, by the inverse Fourier transform of \eqref{eq:FT_w}, one can show that $a(k,t) = \overline{a}(-k,t)$, and hence by symmetry we restrict the solution domain to $k \geq0$. This restricted wavenumber domain is then truncated beyond some $k=k_{\mathrm{end}}$, defining the subdomain $k\in [0,k_{\mathrm{end}}]$, which is discretised using a mesh of $N$ ordered collocation points $k_i$ for $i=0,1,\ldots,N-1$, such that 
\begin{equation}
    0 = k_0 < k_1 < \ldots < k_{N -1} = k_{\text{end}}  .
\end{equation}
The governing equations (\ref{eq:DAE_system_1})--(\ref{eq:DAE_system_3}) are discretised in $k$ by writing $a_i(t) = a(k_i,t)$ and $b_i(t) = b(k_i,t)$ to yield the following system of $2N+1$ ordinary differential-algebraic equations:
\begin{align}
    \diff{a_i}{t} &= b_i,\\
    \diff{b_i}{t} &= -2\mu k_i^2 b_i -  |k_i| \left(k_i^4+Fk_i^2+1\right)a_i,\\
    0 &= 1 - \frac{1}{2}\sum_{j=0}^{N-2} \left( k_{j+1}-k_{j}\right) \left(k_j^2 |a_j|^2 + k_{j+1}^2 |a_{j+1}|^2 \right)  , \label{eq:discrete_length_constraint}
\end{align}
for the $2N+1$ unknowns $a_i, b_i, F$, for $i=0,1,\ldots,N-1$, where \eqref{eq:discrete_length_constraint} has been obtained by applying the trapezoidal rule to the length constraint (\ref{eq:DAE_system_3}). Once equipped with appropriate initial conditions (see Section~\ref{sec:InitialConds}), this system is then passed to GEKKO \cite{beal2018gekko} for numerical integration in time. 
\subsection{Initial conditions}\label{sec:InitialConds}
When equipping (\ref{eq:DAE_system_1})--(\ref{eq:DAE_system_3}) with a set of initial conditions, care must be taken to ensure that any choice is consistent not only with the length constraint \eqref{eq:DAE_system_3}, but also the additional constraints~\eqref{eq:firstderivative}--\eqref{eq:secondderivative} arising from its first and second derivatives.

To obtain a consistent set of initial conditions, an initial profile for $a(k,0)$ is chosen, with an amplitude that has been normalised so that the constraint \eqref{eq:DAE_system_3} is satisfied. In general, the initial velocity $b(k,t)$ must then be prescribed so that \eqref{eq:firstderivative} is satisfied subject to the choice of $a(k,0)$. It is noted, however, for the case in which the system is released from rest $b(k,0)\equiv 0$, the constraint (\ref{eq:firstderivative}) is satisfied for any $a(k,0)$. Since this simplifies the numerical method, we proceed by taking $b(k,0) \equiv 0$ in all of our numerical simulations. Having specified $a(k,0)$ and $b(k,0)$, a corresponding initial value $F(0)$ for the axial tension must then be calculated using \eqref{eq:F_formula}, which in turn ensures that the final constraint, \eqref{eq:secondderivative}, will also be satisfied.

In the numerical results presented below, we consider two sets of initial profiles for the amplitude, $a$, corresponding to Gaussian and rectangular profiles as a function of wavenumber. In each case, the profile is characterised by a parameter crudely dictating its centre $(\alpha >0)$ and width ($\beta>0$).

The Gaussian initial condition takes the form
\begin{equation}\label{eq:init_cond_Gaussian}
    a(k,0) = A_1 \left(\e^{ -\frac{(k-\alpha)^2} {2\beta^2} } + \e^{ - \frac{(k+\alpha)^2} {2\beta^2} } \right) 
    \qquad \text{where} \qquad
    {A_1}^2 =   \frac{2}{\beta^3 \sqrt{\pi}} \left( 1 + 2\frac{ \alpha^2}{\beta^2}  +  \e^{-\frac{\alpha^2}{\beta^2}}  \right) ^{-1},
\end{equation} 
which corresponds to the physical (dimensionless) initial profile
\begin{equation}\label{eq:w_gauss}
    w(x,0)= 2A_1\beta {\sqrt{2L}} \cos(\alpha x) \, \e^{-\frac{\beta^2 x^2}{2}}.
\end{equation}

The rectangular initial condition takes the form (with the additional requirement that $\beta<\alpha$)
\begin{equation}\label{eq:init_cond_rectangular}
    a(k,0) = A_2 \left[  \rect\left(\frac{k-\alpha}{2 \beta}\right)  +\rect\left(\frac{k+\alpha}{2 \beta}\right) \right]
    \qquad \text{where} \qquad
    {A_2}^2 = \frac{3}{2 \beta \left(\beta^2 +3 \alpha^2\right)},
\end{equation}
which corresponds to the physical (dimensionless) initial profile
\begin{equation}\label{eq:w_rect}
    w(x,0)= A_2 \beta   \sqrt{\frac{L}{\pi}} \cos (\alpha x) \, \sinc \left(\frac{\beta x}{\pi} \right) .
\end{equation}

\section{Numerical observations}\label{sec:results}
\subsection{Spectral profiles}
\begin{figure}
    \centering
    \includegraphics[width=\textwidth]{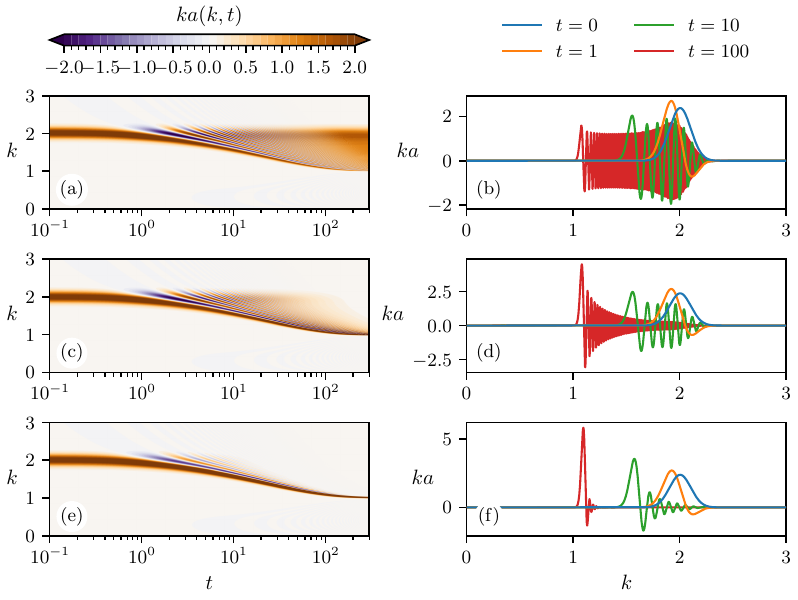}
    \caption{Numerical solutions to mode amplitude equations~\eqref{eq:DAE_system_1}--\eqref{eq:DAE_system_3}, for Gaussian initial data in form of \eqref{eq:init_cond_Gaussian} with $\alpha=2$ and $\beta=1/10$. Left: contour plot of $ka(k,t)$, the length accommodated across wavenumbers $k$ as a function of time $t$.  Right: $ka$ as a function of $k$ at times $t = 0$, $1$, $10$ and $100$. The dissipation coefficient $\mu$ is set to \{0, 0.005, 0.05\} top to bottom. 
    }
    \label{fig:Gaussian_length_contribution_panels}
\end{figure}

\begin{figure}
    \centering
    \includegraphics[width=\textwidth]{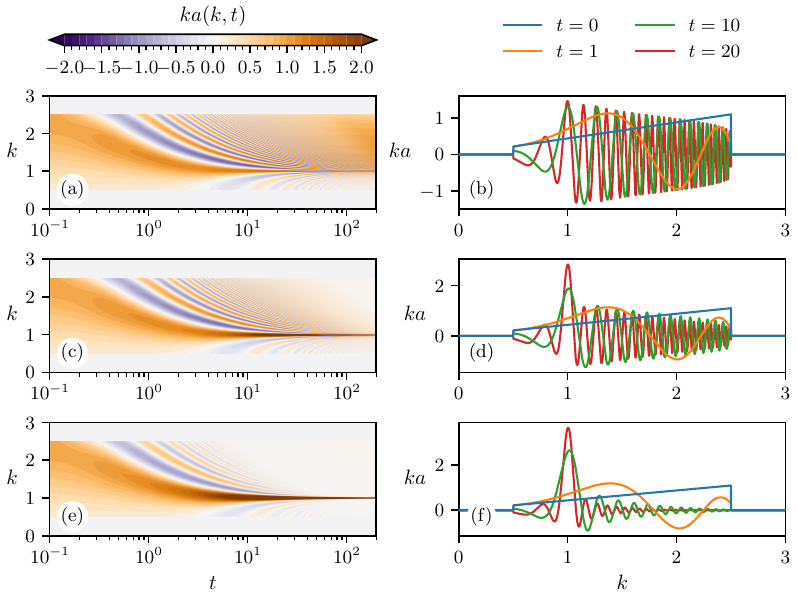}
    \caption{Numerical solutions to mode amplitude equations~\eqref{eq:DAE_system_1}--\eqref{eq:DAE_system_3}, for rectangular initial data in form of \eqref{eq:init_cond_rectangular} with $\alpha=1.5$ and $\beta=1$. Left: contour plot of $ka(k,t)$, the length accommodated across wavenumbers $k$ as a function of time $t$.  Right: $ka$ as a function of $k$ at times $t = 0$, $1$, $10$ and $100$. The dissipation coefficient $\mu$ is set to \{0, 0.005, 0.05\} top to bottom.
    }
    \label{fig:Rectangular_length_contribution_panels}
\end{figure}

We solve the governing equations presented in Section~\ref{sec:fourier} via the numerical procedure outlined in Section~\ref{sec:gekko} and present solutions as a function of wavenumber and time for the Gaussian initial conditions~\eqref{eq:init_cond_Gaussian} (see Figure~\ref{fig:Gaussian_length_contribution_panels}) and rectangular initial conditions~\eqref{eq:init_cond_rectangular}  (see Figure~\ref{fig:Rectangular_length_contribution_panels}). The quantity $|ka(k,t)|^2$ describes the length accommodated by each wavenumber $k$ as a function of time, and therefore we have made the decision to scale amplitudes $a$ with a factor of $k$ before plotting.  For the Gaussian initial condition, the wavepacket first spreads in $k$-space, transferring length to lower values of $k$ (longer wavelengths). In the absence of dissipation ($\mu=0$), a broad spectrum of modes remain excited at late times (Figure~\ref{fig:Gaussian_length_contribution_panels}(a)), whilst when dissipation is present ($\mu \neq 0$) (Figure~\ref{fig:Gaussian_length_contribution_panels}(c), (e)), the wavepacket narrows at a rate that increases with increasing $\mu$ and the peak moves towards $k=1$. Using rectangular initial condition leads to similar overall wrinkle dynamics (Figure~\ref{fig:Rectangular_length_contribution_panels}). A broader initial excitation across $k \in [0.5,2.5]$ leads to wider wavepackets, which nonetheless show a peak that converges towards $k = 1$ and a focusing that occurs faster for larger $\mu$, as expected. In both cases, we see that eventually the solutions converge to the late-time steady-state equilibrium $k = 1$, as predicted from the energetic arguments made in Section \ref{sec:EnergeticConsiderations}. Interestingly, in all cases the compressive force decays to the steady-state value predicted in Section~\ref{sec:EnergeticConsiderations}, even though this steady state cannot be achieved in the absence of dissipation.

\begin{figure}
    \centering
    \includegraphics[width=\textwidth]{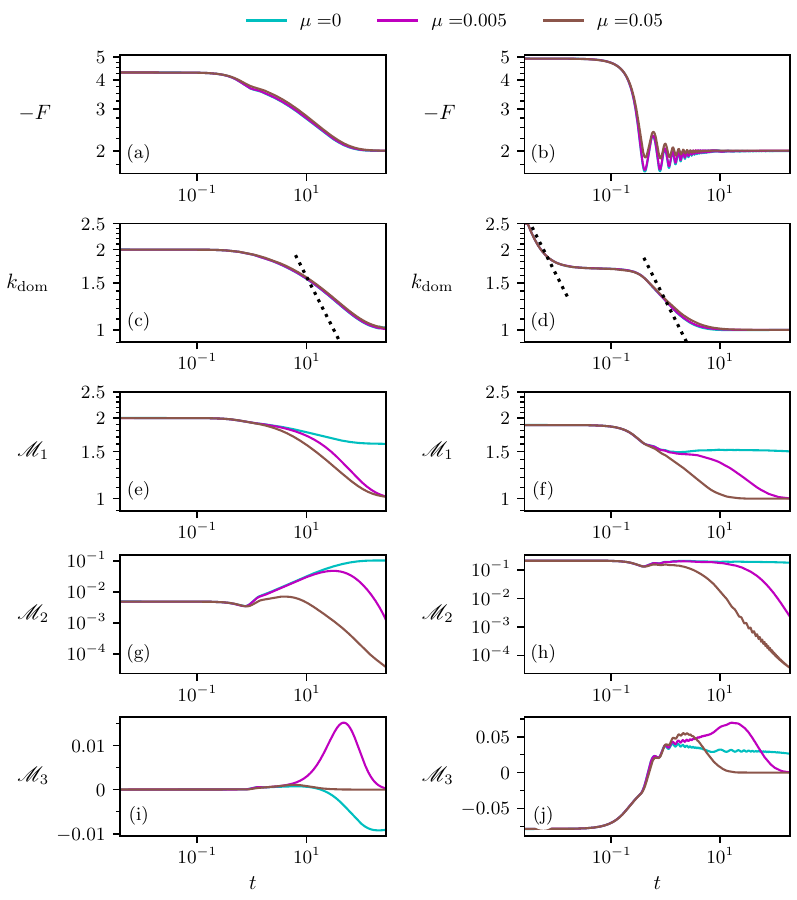}
    \caption{The magnitude of the compressive force, $-F(t)$, the dominant wavenumber $k_{\mathrm{dom}}$, and the moments $\mathscr{M}_i$ for $i=1,2,3$ for the Gaussian (left) and rectangular (right) initial conditions. Dashed lines are $\sim t^{-2/5}$.
    }
    \label{fig:metrics}
\end{figure}

Examining Figures~\ref{fig:Gaussian_length_contribution_panels} and \ref{fig:Rectangular_length_contribution_panels}, it can be observed that the profiles for $ka$ become more oscillatory in $k$ as time increases; an explanation for this is given in Section~\ref{sec:wavenumber_metrics}. Despite these oscillations, it appears that the dynamics of the peak are effectively independent of the dissipation coefficient, $\mu$, even though the spectral profiles exhibit significant variability for different $\mu$.  We illustrate this further by defining a dominant wavenumber 
\begin{equation}\label{eq:dom_k}
    \kdom(t)=\argmax_{k>0}\Big{(}|a(k,t)|\Big{)},
\end{equation}
which is plotted in Figure~\ref{fig:metrics}(c) as a function of time. It is observed that the dominant wavenumber is effectively independent of $\mu$ for a given initial condition.  The Gaussian initial condition considered here excites a narrow range of $k$-values close to $k=2$, corresponding to sufficiently large wavelengths where buoyancy is already significant at $t = 0$ and an inertia--bending regime does not occur. In this case, the early-time dynamics for $\kdom$ do not align with the $t^{-2/5}$ trend suggested by~\cite{okiely2020impact} for regimes dominated by inertia and bending. 
By contrast, the early-time dynamics for the rectangular initial profile do appear to align with the $t^{-2/5}$ trend suggested by~\cite{okiely2020impact}.
An explanation for this is that the rectangular initial condition allocates significant length to $k$ values near $k=2.5$; these short-wavelength modes grow rapidly initially, allowing the $t^{-2/5}$ trend associated with inertia--bending to emerge briefly, before plateauing and evolving more slowly through a combination of inertia, bending and buoyancy. 

The different early-time trends for different initial conditions can be further justified by consulting the definition \eqref{eq:dom_k} of the dominant wavenumber. This definition selects only the single wavenumber associated with the largest contribution to the global amplitude without consideration of the relative importance of contributions from other wavenumbers. Consequently, for smooth Gaussian initial conditions, any mode which is not the initial maximum of $a$ requires significant time before its amplitude is able to overcome that of the central peak, and this is exacerbated by imposing particularly narrow initial profiles. By contrast, for rectangular initial conditions, which have uniform amplitudes, dominant mode values can, from the outset, emerge from the initial profile.

\subsection{Wavenumber distribution metrics}\label{sec:wavenumber_metrics}
 Whilst the dominant wavenumber provides a useful metric for analysing wrinkle dynamics, it fails to take into account the non-negligible contributions from the remaining Fourier modes. For this reason, in this section we introduce alternative (statistical) metrics that may be used to further analyse the numerical results presented above by considering the central moments (namely the mean, variance and skewness) that arise through treating our solution spectra as density functions associated with the length allocated to each wrinkling mode. That is, we define $n$th-order raw moments
\begin{align}
  \left\langle k^n \right\rangle = \int_{0}^{\infty} k^n \, P(k,t)\, \mathrm{d}k,
\end{align}
where $P(k,t) = k^2 |a(k,t)|^2$ denotes the length accommodated by a wavenumber $k$ with amplitude $a$, and $\int_0^{\infty} P(k,t)\ \mathrm{d} k = 1$ by the length constraint (\ref{eq:DimensionlessLengthConstraint}), so that $P$ takes the form of a standard probability density distribution.  The mean, variance and skewness can then be defined as
\begin{equation}
\mathscr{M}_1(t) = \langle k \rangle, \qquad \mathscr{M}_2(t) = \Big\langle\big(k-\langle k\rangle\big)^2\Big\rangle, \qquad \mathscr{M}_3(t) = \Big\langle\big(k-\langle k\rangle\big)^3\Big\rangle.
\end{equation}

Unlike the dominant wavenumber, the mean value $\mathscr{M}_1$ is dependent on the dissipation coefficient $\mu$ (see Figure~\ref{fig:metrics}(e), (f)). For $\mu = 0$, the mean does not appear to be converging to $\mathscr{M}_1=1$ even at later times. This is consistent with the late-time spread observed in Figures~\ref{fig:Gaussian_length_contribution_panels}(a), (b) and \ref{fig:Rectangular_length_contribution_panels}(a), (b). For $\mu = 0.05$, once the early-time transients have passed, the mean $\mathscr{M}_1$ appears to track the peak $\kdom$ closely; this is consistent with the narrow focusing observed in Figures~\ref{fig:Gaussian_length_contribution_panels}(e), (f) and \ref{fig:Rectangular_length_contribution_panels}(e), (f). Solutions corresponding to the intermediate value $\mu = 0.005$ (plotted in Figures~\ref{fig:Gaussian_length_contribution_panels}(c), (d) and \ref{fig:Rectangular_length_contribution_panels}(c), (d)) show that the value of $\mathscr{M}_1$ does evolve towards 1, but does so more slowly than $\kdom(t)$. 

Similar to the mean, the variance, $\mathscr{M}_2$, also exhibits significant variability with respect to changes in the dissipation parameter, $\mu$ (see Figure~\ref{fig:metrics}(g), (h)). Specifically, when $\mu =0$ the variance plateaus to a constant (non-zero) value, whereas when $\mu \neq 0$ the variance approaches zero at late times, with a rate that increases with an increase in the dissipation coefficient. For a uniform initial condition there is a bias in potential energy towards higher wavenumbers (as is confirmed by (\ref{eq:AppendixEnergyExpressions})) owing to the fact that high-frequency wrinkles store more bending energy, which scales with the square of the wrinkle curvature. Minimising the potential energy therefore relies upon transferring potential energy from higher-to-lower wavenumbers, which by the length constraint modifies the amplitude of each mode over time, and hence populates the modes with kinetic energy. When populated with kinetic energy, inertial effects then provide a mechanism for the broadening of the wavepacket, as observed in Figure~\ref{fig:metrics}(g), (h)) above, before dissipative effects overcome inertia leading to narrowing again.

To gain further insight, we consider the evolution equation~\eqref{eq:ak_equation} through the lens of a damped harmonic oscillator and treat the compressive force $F<-2$ as though it were fixed, with $\mu \ll 1$, which yields approximate analytical solutions
\begin{equation}\label{eq:oscillator}
a(k,t) = A(k) \mathrm{e}^{-\mu |k|^2 t} \cos\left(t \, \sqrt{|k|(k^4+Fk^2+1)}\right),
\end{equation}
from which we can deduce that each wrinkle mode oscillates in time at a frequency that increases with both $|k|$ and the distance from the roots $k^2$ of $k^4+Fk^2+1=0$, which we interpret as quasi-steady ``preferred'' wavenumbers. The oscillations in time transform to vertical velocities in physical space, and are penalised by the viscous dissipation term $2 \mu {\partial a}/{\partial t}$ in (\ref{eq:ak_equation}), which again increases with $|k|$ and the distance from the quasi-steady wavenumbers, and is also proportional to $\mu$. In agreement with our numerical results, (\ref{eq:oscillator}) suggests that the peak $\kdom$ evolves in tandem with the compressive force $F(t)$, which is plotted in Figure~\ref{fig:metrics}(a), (b) and decays towards $F = -2$ as $\kdom \to 1$, as expected. Viscous damping focuses the wrinkle spectrum around this peak; this explains the trends for $\mathscr{M}_1$ for $\mu\neq0$ in Figure~\ref{fig:metrics}(e), (f) and also suggests that following initial transients the variance should decay in time at a rate that depends on $\mu$, as we have observed in Figure~\ref{fig:metrics}(g), (h). Equation (\ref{eq:oscillator}) also helps to inform the dynamics of the skewness (plotted in Figure~\ref{fig:metrics}(i), (j)), where we see that  the focusing of the wavepacket should be asymmetric about $\kdom$, with large-$k$ modes dissipating more strongly, so that a positive skew grows over a timescale that is faster for smaller $\mu$, and subsequently decays as the spectrum converges on the equilibrium configuration.

One feature that remains to be discussed are the oscillations in $F$ for $t = O(1)$ for rectangular initial conditions.  We note that these are not simply a consequence of the piecewise initial conditions; the oscillations remain if the initial condition is smoothed by considering a set of $\tanh$ profiles rather than a rectangle. For the case of rectangular initial conditions, the compressive force decays rapidly and overshoots the $F = -2$ target.  We interpret this as coming from high-amplitude modes that initially accelerate to grow and then, with inertia, continue to grow beyond the point where they are energetically favourable.  This leads to an over-accommodation of excess length and a rapid decay in compressive force that is out of sync with $\kdom$.  After $F$ increases above $-2$, there is no longer a real-valued ``preferred'' wavenumber associated with roots of $k^4 + Fk^2 + 1$, so the system adjusts and the process restarts.  Increasing viscous damping decreases this overshooting effect, as expected, and a signature of the overshoot oscillations can be seen in the mean and variance. 

\subsection{Energy}
\begin{figure}
    \centering
    \includegraphics[width=\textwidth]{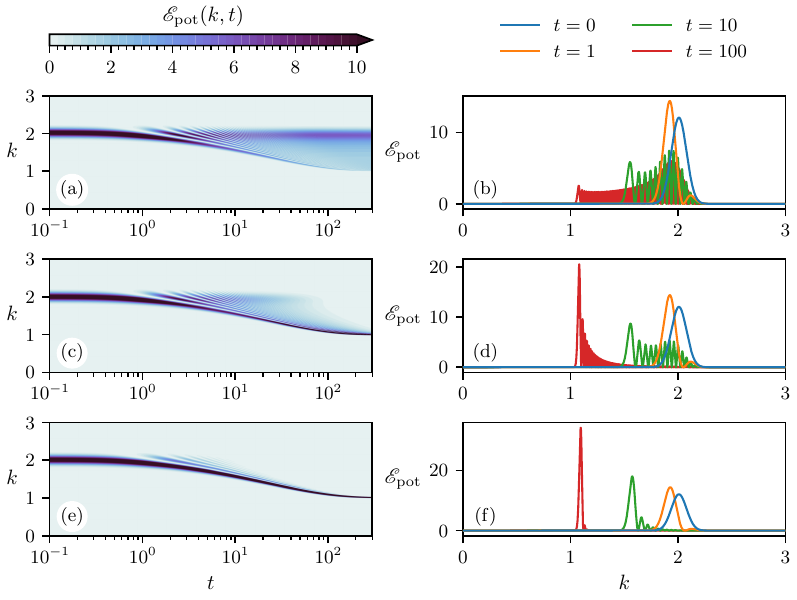}
    \caption{Left: Potential energy density contours. Right: Energy density at chosen times. The dissipation values $\mu$ are \{0, 0.005, 0.05\} top to bottom.
    Gaussian initial data in form of \eqref{eq:init_cond_Gaussian} with $\alpha=2$ and $\beta=1/10$.
    }
    \label{fig:Gaussian_pe_density_panels}
\end{figure}

\begin{figure}
    \centering
    \includegraphics[width=\textwidth]{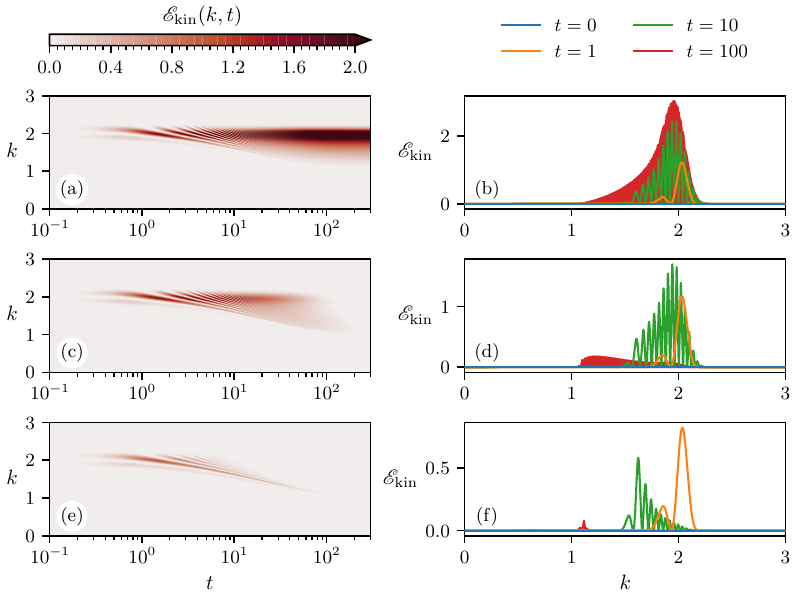}
    \caption{Left: Kinetic energy density contours. Right: Energy density at chosen times. The dissipation values $\mu$ are \{0, 0.005, 0.05\} top to bottom.
    Gaussian initial data in form of \eqref{eq:init_cond_Gaussian} with $\alpha=2$ and $\beta=1/10$.
    }
    \label{fig:Gaussian_ke_density_panels}
\end{figure}

We plot the potential and kinetic energy distributions for a Gaussian initial condition in Figures~\ref{fig:Gaussian_pe_density_panels} and \ref{fig:Gaussian_ke_density_panels} respectively; these correspond to the spectrum illustrated in Figure~\ref{fig:Gaussian_length_contribution_panels}.  The potential energy starts concentrated around $k=2$ and the kinetic energy starts at zero everywhere by the initial conditions.  As time evolves, the potential energy profile broadens as lower modes are populated with potential energy, and the kinetic energy increases from zero everywhere; if the system has dissipation it later decreases back toward zero everywhere.  Energy is continuously transferred between kinetic and potential until at late time when the potential energy is concentrated around the equilibrium solution $k=1$.  As expected from {Figure~\ref{fig:Gaussian_length_contribution_panels}}, both the potential and kinetic energy profiles become more oscillatory in $k$ as time increases (see Figure~\ref{fig:Gaussian_length_contribution_panels}), and the oscillations are wider at smaller $k$.

The position of the peak in potential energy moves gradually, and is independent of dissipation, consistent with the amplitude peak $\kdom$ in Figure~\ref{fig:metrics}(c).  Nonetheless, the profile is influenced somewhat by dissipation: in the absence of dissipation some potential energy remains around the position of the initial peak, while increasing the dissipation parameter from zero narrows the peak of the profile, consistent with the trend in $\mathscr{M}_2$ illustrated in Figure~\ref{fig:metrics}(g).  By contrast, the position of the peak in kinetic energy as a function of time depends strongly on the dissipation parameter.  When $\mu = 0.05$, the peak in kinetic energy closely tracks the peak in the potential energy, because modes away from this peak are dissipated strongly.  However, as $\mu$ decreases the movement of the peak in kinetic energy is delayed, and when $\mu = 0$ it remains close to the initial peak.  We also note that while the kinetic energy peak is close to $k=2$, its profile has a clear tail to the left, whereas after it has travelled toward $k=1$ it has a tail to the right.

\begin{figure}
    \centering
    \includegraphics[width=\textwidth]{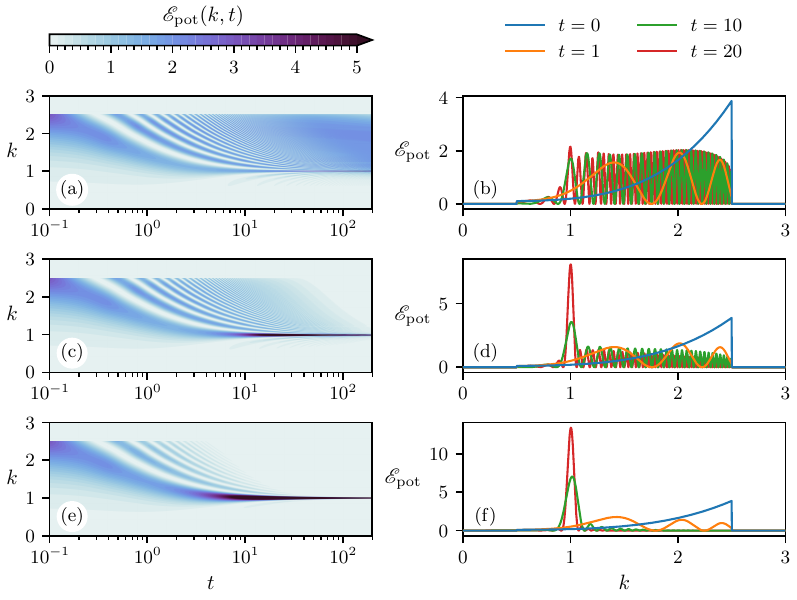}
    \caption{Left: Potential energy density contours. Right: Energy density at chosen times. The dissipation values $\mu$ are \{0, 0.005, 0.05\} top to bottom.
    Rectangular initial data in form of \eqref{eq:init_cond_rectangular} with $\alpha=1.5$ and $\beta=1$.
    }
    \label{fig:Rectangular_pe_density_panels}
\end{figure}

\begin{figure}
    \centering
    \includegraphics[width=\textwidth]{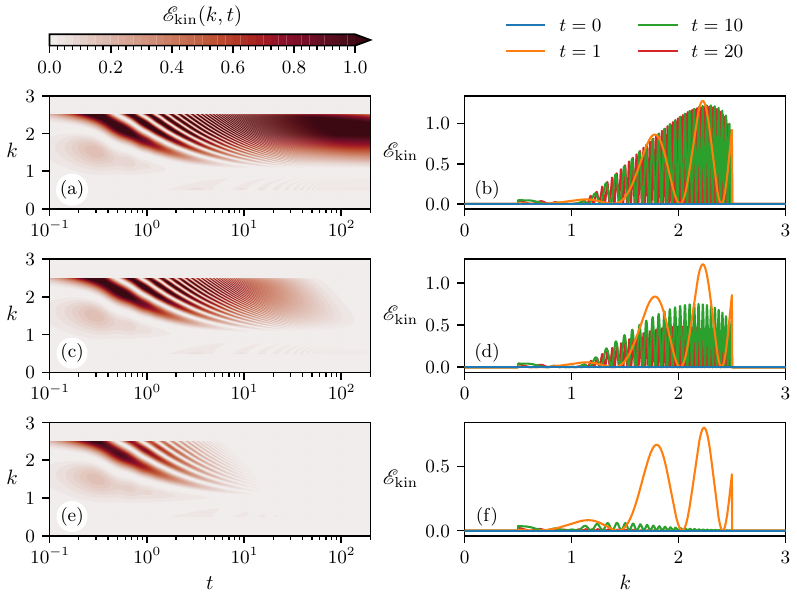}
    \caption{Left: Kinetic energy density contours. Right: Energy density at chosen times. The dissipation values $\mu$ are \{0, 0.005, 0.05\} top to bottom.
    Rectangular initial data in form of \eqref{eq:init_cond_rectangular} with $\alpha=1.5$ and $\beta=1$.
    }
    \label{fig:Rectangular_ke_density_panels}
\end{figure}

Potential and kinetic energy distributions for a rectangular initial condition are plotted in Figures~\ref{fig:Rectangular_pe_density_panels} and \ref{fig:Rectangular_ke_density_panels} respectively; these correspond to the spectrum illustrated in Figure~\ref{fig:Rectangular_length_contribution_panels}.  For the most part, we observe the same qualitative trends as for the Gaussian initial condition. However, since for a rectangular initial condition the higher modes have significantly larger potential energy than for the Gaussian, the transfer of potential energy into kinetic energy and the subsequent dissipation are both more extreme and so the qualitative changes outlined above occur on a much faster timescale for the rectangular initial condition.  Some additional differences in the shape of the potential energy can be observed: the initial potential energy has a monotonic $k^4+1$ shape in the excited $k$-range, but develops rapidly into a wavetrain (a wavepacket with approximately constant envelope height) for all dissipation values, while the corresponding kinetic energy rapidly develops a local peak.  The peak then evolves in a similar fashion to the peak for Gaussian initial conditions.

\begin{figure}
    \centering
    \includegraphics[width=1.0\textwidth]{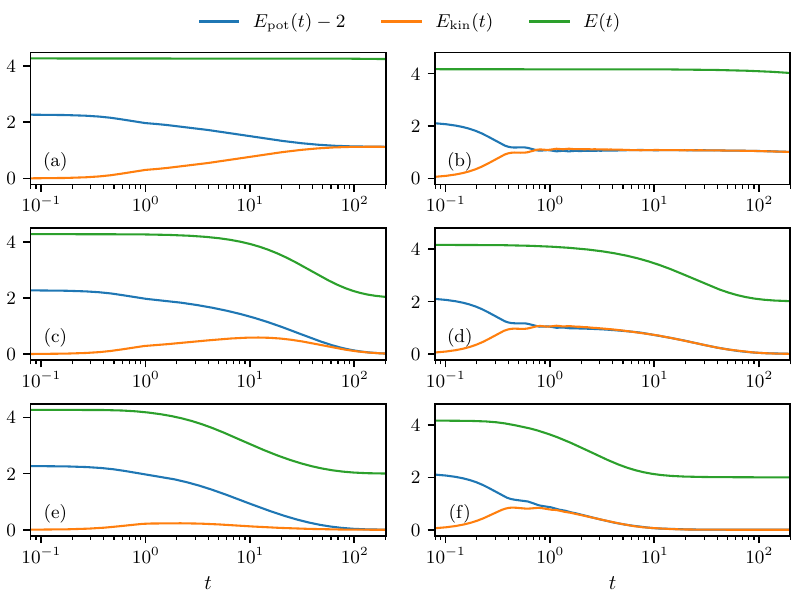}
    \caption{Excess potential, kinetic, and total energies as a function of time, where the excess potential is measured relative to the minimum potential of $2$ in steady state. Left: Gaussian initial condition, right: Rectangular initial condition. The dissipation values $\mu$ are \{0, 0.005, 0.05\} top to bottom. 
    }
    \label{fig:energy_and_equipartition}
\end{figure}

We can gain some additional insight by plotting total energies in Figure~\ref{fig:energy_and_equipartition}; that is, integrating the quantities plotted in Figure~\ref{fig:Gaussian_pe_density_panels}---\ref{fig:Rectangular_ke_density_panels} in $k$.  From these, it is clear that the systems considered in our numerical simulations all start with no kinetic energy (as imposed by the initial conditions) and some potential energy.  For the purpose of these plots, we measure potential as an excess above the value of $2$ that corresponds to the energy-minimising steady state.  We observe that excess potential is transferred into kinetic energy until equipartition is reached, with or without dissipation.  With dissipation, this corresponds to complete removal of both kinetic and excess potential energy, while in the absence of dissipation this corresponds to equal sharing of the initial excess between potential and kinetic forms.  In all cases, the energy transfer occurs concurrently with the drop in compressive force $F$ and the transfer of energy and length from high to low wavenumbers.  In the case of rectangular initial conditions, there is a slight overshoot, consistent with the overshoot of $F$ in Figure~\ref{fig:metrics}(b).

\section{Conclusions}\label{sec:conclusions}
When the ends of a thin elastic sheet are pushed together, the elastic sheet typically buckles out-of-plane, accommodating its end-to-end displacement while preserving its arclength.  It is well-known that for small strains, the resulting buckling pattern takes a sinusoidal form, with a wavelength determined by competition between bending stiffness, which resists curvature, and restoring stiffnesses, which resist displacement.  Examples of resistive stiffnesses include the buoyancy or elastic stiffness of a liquid or soft solid substrate, inertia, and elastic contributions from the sheet itself associated with transverse tension or curvature.

In this paper, we have presented a study of wrinkles in a uniaxially compressed floating sheet, evolving from an initial out-of-plane perturbation, through early-time inertia-dominated growth, to late-time gravity-controlled equilibrium.  The wavelength of the initial disturbance strongly affects the early-time dynamics: for small-wavelength perturbations, the only restoring force comes from inertia and a scaling law predicted elsewhere in the literature can be observed, while perturbations at relatively long wavelengths are affected by gravity even at early times.

For an infinitely long sheet on an infinite low-viscosity bath, the wrinkle profile in the thin elastic sheet can be decomposed into Fourier modes, whose amplitudes evolve in time under the combined action of compression, inertia and gravity, and are coupled by a length constraint which takes the form of a nonlinear algebraic equation.  The presence of this \emph{spectrum} of wrinkling modes is essential; a single wrinkling mode whose wavenumber evolves in time cannot satisfy the governing equations for the system.  Upon numerical solution of the amplitude equations, our model captures simultaneous coarsening of the wrinkles and collapse of the compressive stress in the system: the peak of the wrinkle spectrum shifts as wrinkles of different wavelengths grow and decay.  At late time, the system reaches an energetically-favourable steady state at a single buoyancy-selected wavenumber, provided that there is a mechanism for excess energy to be dissipated from the system.  We also observed equipartition of the global energies in the sheet at late time.

Dissipation was included in the model in the form of a small-but-finite viscous contribution from the bath of liquid that the sheet floats on.  Here, we adapted an existing analysis of surface waves in a fluid, invoking a Helmholtz decomposition to resolve the flow in the underlying fluid bath and hence derive a viscous drag on the sheet which acts opposition to its vertical velocity.  We found that viscous dissipation narrows the spectrum of excited wrinkling modes: each mode oscillates with a frequency that increases with distance from the spectrum's peak, and so modes in the tail of the amplitude distribution are affected disproportionately by drag.  It is therefore surprising that our results also suggest that both the position of the peak of the spectrum and the compressive force in the sheet evolve along a path that is independent of the strength of dissipation.  We hope this intriguing finding will stimulate further study of dynamic wrinkle coarsening.

Focusing on energy, we observed that evolution of the wrinkle spectrum is closely tied to transfer of energy, as expected.  This transfer occurs both between potential and kinetic energy and between modes.  Any initial amplitude which is not a narrow peak at $k=1$ has an excess potential energy, which induces motion and transfer to kinetic energy.  With dissipation, this kinetic energy is continually removed and therefore so is excess potential until the steady state is reached.  Without dissipation, the initial excess energy is shared between kinetic and potential and so a steady state is never reached.

\begin{figure}
    \centering
    \includegraphics[width=\textwidth]{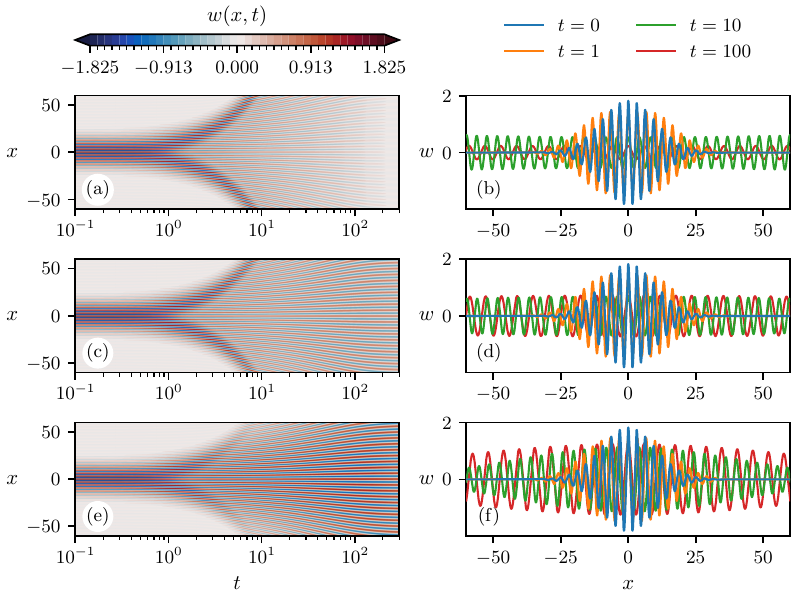}
    \caption{Left: wrinkle contours. Right: Wrinkle profiles in space at chosen times. The dissipation values $\mu$ are \{0, 0.005, 0.05\} top to bottom.
    Gaussian initial data in form of \eqref{eq:init_cond_Gaussian} with $\alpha=2$ and $\beta=1/10$.
    }
    \label{fig:Gaussian_wrinkle_profiles}
\end{figure}

\begin{figure}
    \centering
    \includegraphics[width=\textwidth]{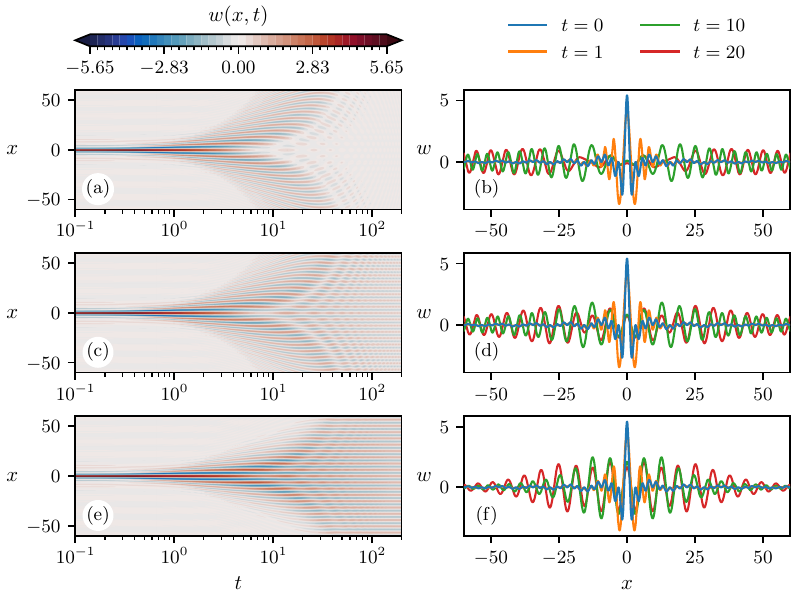}
    \caption{Left: wrinkle contours. Right: Wrinkle profiles in space at chosen times. The dissipation values $\mu$ are \{0, 0.005, 0.05\} top to bottom.
    Rectangular initial data in form of \eqref{eq:init_cond_rectangular} with $\alpha=1.5$ and $\beta=1$.
    }
    \label{fig:Rectangular_wrinkle_profiles}
\end{figure}

Finally, we can reconstruct the wrinkle profile in physical space from the amplitude $a(k,t)$.  We do this in Figures \ref{fig:Gaussian_wrinkle_profiles} and \ref{fig:Rectangular_wrinkle_profiles}, which show the spatial wrinkles and their evolution through time for the Gaussian and rectangular initial conditions, respectively. Both sets of results nicely demonstrate the effects discussed in terms of energy and modes, above. Both sets of wrinkles visibly coarsen throughout the dynamics resulting in steady parallel oscillations with a period greater than the initial configuration. The formation of an equilibrium is demonstrated in each of the contour panels (a), (c), and (e), through the broadening of the wrinkle distribution but additionally highlights that the formation of the steady state requires the amplitude of $w$ to increase.

\section*{Acknowledgements}
The research leading to these results has received funding from the Australian Research Council under award number DP230100406 (DJN), from Research Ireland under the Frontiers for the Future Programme, grant no. 21/FFP-P/10160 (DOK) and from the Simons Foundation (DOK).  
This research was carried as a result of the authors' participation in an Isaac Newton Institute satellite programme ``The mathematics of multiphase flows with applications'', organised by Prof.\ Mark Blyth, Prof. Jose Manuel Gordillo, Prof. Alexander Korobkin, and hosted by the University of East Anglia (supported by EPSRC grant EP/V521929/1).

\color{black}

\bibliographystyle{RS}
\bibliography{main}

\appendix
\numberwithin{equation}{section} 
\section{Derivation of the fluid-mechanical equations}\label{sec:AppendixDias}
\subsection{Governing equations}
The fluid flow is governed by the Navier--Stokes equations, which in the absence of nonlinear fluid inertia are given by
\begin{align}
    \nabla^* \cdot \bm{v}^* &=0, \label{eq:DimensionalFluidMassConservation} \\
    \rho^* \frac{\partial \bm{v}^*}{\partial t^*} &= -\nabla^* p^* +\mu^* {\nabla^*}^2 \bm{v}^* - \rho^*g^* \bm{k},  \label{eq:DimensionalFluidMomentumConservation}
\end{align}
where $\bm{v}^* = v_x^*\bm{i}+v_z^*\bm{k}$ is the fluid velocity field, $t^*$ is time, $p^*$ is the hydrodynamic pressure, $\rho^*$ is the fluid density, $\mu^*$ is the dynamic viscosity, $g^*$ is the force due to gravity and $\nabla^*$ is the spatially two-dimensional gradient operator. 

There are two boundary conditions to be imposed at the interface between the beam and the fluid. The first is the usual kinematic condition
\begin{equation}\label{eq:ViscoInertialKinematicCond}
    v_z^* = \frac{\partial w^*}{\partial t^*}+v_x^*\frac{\partial w^*}{\partial x^*}, \qquad \text{at} \qquad z^* = w^*,
\end{equation}
The second condition enforces stress continuity at the fluid-beam interface, and is given by 
\begin{equation}\label{eq:DimensionalStressBalance}
    (-p^*I_2+2\mu^*e)\cdot \bm{n} = -p^*_{\mathrm{ext}}\bm{n} \qquad \text{at} \qquad z^* = w^*,
\end{equation}
where $I_2$ is the $2 \times 2$ identity matrix, $e$ is the rate-of-strain tensor, $p_{\mathrm{ext}}^*$ is the external pressure (normal force per unit area) acting on the beam and $\bm{n}$ is the unit normal, which by convention is positive when oriented outwards relative to the fluid. 

The small-amplitude displacement regime considered here means that the unit normal points predominantly in the $\bm{k}$ direction and hence the normal and tangential components of the stress balance (\ref{eq:DimensionalStressBalance}) are given by
\begin{equation}\label{eq:DimensionalNormalStressBalance}
     -p^* + 2\mu^* \pdiff{v^*_z}{z^*} = -p^*_{\mathrm{ext}} \qquad \text{at} \qquad z^* = w^*,
\end{equation}
and 
\begin{equation}\label{eq:DimensionalTangentialStressBalance}
    \mu^* \left(\pdiff{v_x^*}{z^*} + \pdiff{v_x^*}{z^*}\right) = 0 \qquad \text{at} \qquad z^* = w^*.
\end{equation}
Our analysis also requires prescription of far-field conditions on the fluid velocity, however, we defer this for the moment. 
\subsection{Potential flow}\label{sec:PotentialFlow}
We now resolve the fluid velocity field into irrotational and solenoidal parts using the Helmholtz decomposition \cite{dias2008theory}
\begin{equation}\label{eq:HelmholtzDecomp}
    \bm{v}^* = \nabla^* \phi^* + \nabla^* \times \bm{\psi}^* = \left(\pdiff{\phi^*}{x^*}-\pdiff{\psi^*}{z^*}\right)\bm{i} + \left(\pdiff{\phi^*}{z^*}+ \pdiff{\psi^*}{x^*}\right)\bm{k},
\end{equation}
where $\bm{\psi} = \psi^*\bm{j}$. Substituting (\ref{eq:HelmholtzDecomp}) into (\ref{eq:DimensionalFluidMassConservation}), it follows that $\phi^*$ satisfies Laplace's equation
\begin{equation}\label{eq:DimensionalLaplacesEquation}
    {\nabla^*}^2 \phi^* = 0.
\end{equation}
Similarly, on substituting (\ref{eq:HelmholtzDecomp}) into (\ref{eq:DimensionalFluidMomentumConservation}) and making use of (\ref{eq:DimensionalLaplacesEquation}), we obtain the following pair of equations
\begin{align}
    \rho^*\pdiff{}{t^*}\left(\pdiff{\phi^*}{x^*}-\pdiff{\psi^*}{z^*}\right) &= -\pdiff{p^*}{x^*} - \mu^*\frac{\partial }{\partial z^*}\left({\nabla^*}^2 \psi^*\right), \label{eq:NS_decomp1} \\
    \rho^*\pdiff{}{t^*}\left(\pdiff{\phi^*}{z^*}+\pdiff{\psi^*}{x^*}\right) &= -\pdiff{p^*}{z^*} + \mu^*\frac{\partial }{\partial x^*}\left({\nabla^*}^2 \psi^*\right)-\rho^*g^*. \label{eq:NS_decomp2}
\end{align}
By differentiating (\ref{eq:NS_decomp1}) in $z^*$ and (\ref{eq:NS_decomp2}) in $x^*$, subtracting and then integrating, one can deduce that $\psi^*$ satisfies
\begin{equation}\label{eq:DimensionalPsiGoverningEquation}
    \mu^* {\nabla^*}^2 \psi^* = \rho^*\pdiff{\psi^* }{t^*},
\end{equation}
where it is noted that by virtue of having integrated in two spatial dimensions one has the freedom to apply a time and/or spatially linear gauge transformation to $\psi^*$, with a corresponding transformation also required to be imposed on the velocity potential $\phi^*$ for consistency with (\ref{eq:HelmholtzDecomp}). Since the resultant fluid velocity $\bm{v}^*$ is invariant to any gauge transformation of this form, we proceed by taking (\ref{eq:DimensionalPsiGoverningEquation}) in isolation. We also note that it is unsurprising that $\psi^*$ satisfies the diffusion equation, with the diffusion coefficient equal to the kinematic viscosity.

By considering equation (\ref{eq:NS_decomp2}), making use of (\ref{eq:DimensionalPsiGoverningEquation}), and then integrating, it is found that
\begin{equation}\label{eq:LinearisedBernoulli}
    \rho^*\pdiff{\phi^*}{t^*} = -p^* - \rho^* g^* z^*,
\end{equation}
where the arbitrary constant has been absorbed into the potential $\phi^*$. In terms of the potentials $\phi^*$ and $\psi^*$ the kinematic equation is given by
\begin{equation}\label{eq:KinematicPotential}
    \pdiff{\phi^*}{z^*}+ \pdiff{\psi^*}{x^*}  =\frac{\partial w^*}{\partial t^*}+\left(\frac{\partial \phi^*}{\partial x^*}-\frac{\partial \psi^*}{\partial z^*}\right)\frac{\partial w^*}{\partial x^*} \qquad \text{at} \qquad z^* = w^*.
\end{equation}
Using (\ref{eq:DimensionalPsiGoverningEquation}), the tangential component of the stress condition (\ref{eq:DimensionalTangentialStressBalance}) is given by 
\begin{equation}
    2\left(\frac{\partial^2 \phi^*}{\partial z^* \partial x^*}-\frac{\partial^2 \psi^*}{\partial {z^*}^2}\right) +\frac{\rho^*}{\mu^*}\pdiff{\psi^*}{t^*} = 0.
\end{equation}
Finally, eliminating the hydrodynamic pressure between (\ref{eq:LinearisedBernoulli}) and the normal-stress balance (\ref{eq:DimensionalNormalStressBalance}) gives to leading order
\begin{equation}\label{eq:ViscoInertialDynamicCond}
      \rho^*\pdiff{\phi^*}{t^*} + \rho^* g^* w^* + 2 \mu^* \left(\frac{\partial^2 \phi^*}{\partial {z^*}^2}+\frac{\partial^2 \psi^*}{\partial z^* \partial x^*}\right) = -p_{\mathrm{ext}}^* \qquad \text{at} \qquad z^* = w^*. 
\end{equation}
When combined with appropriate far-field fluid-velocity conditions, equations (\ref{eq:DimensionalLaplacesEquation}), (\ref{eq:DimensionalPsiGoverningEquation}) and (\ref{eq:KinematicPotential})--(\ref{eq:ViscoInertialDynamicCond}) constitute the fluid-mechanical governing system under consideration.

\section{Energetics}\label{sec:AppendixEnergy}
In this appendix we derive an expression for the energy budget of the system (\ref{eq:ak_equation})--(\ref{eq:a_k_integralconstraint}). We begin by first taking the sum of the product between $ \partial \overline{a}/ \partial t$ and \eqref{eq:ak_equation} and the product between $\partial a / \partial t$ and the conjugate of \eqref{eq:ak_equation} to obtain 
\begin{equation}\label{eq:EnergyBudgetv1}
    \frac{1}{|k|} \left( \pdiff{\overline{a}}{t} \pdiff[2]{a}{t} + \pdiff{a}{t} \pdiff[2]{\overline{a}}{t} \right)
    + 4\mu |k|  \pdiff{a}{t}  \pdiff{\overline{a}}{t}  
    + \left( k^4 + F k^2 +1 \right) \left( \pdiff{\overline{a}}{t} a + \pdiff{a}{t} \overline{a}  \right)  = 0 .
\end{equation}
Using the product rule for differentiation, equation (\ref{eq:EnergyBudgetv1}) may be written as 
\begin{equation}\label{eq:EnergyBudgetv2}
    \frac{1}{2|k|} \pdiff{}{t} \bigg|\pdiff{a}{t}\bigg|^2 
    + 2\mu |k| \bigg|\pdiff{a}{t}\bigg|^2
    + \frac{1}{2} \left( k^4 + F k^2 +1 \right)  \pdiff{}{t} |a|^2 = 0 .
\end{equation}
Integrating (\ref{eq:EnergyBudgetv2}) over all wavenumbers $k \in \mathbb{R}$, exchanging the order of time derivatives with the integrals, and rearranging, it follows that
\begin{equation}\label{eq:EnergyBudgetv3}
    \diff{}{t}\left[\int^{\infty}_{-\infty} \frac{1}{2|k|} \bigg|\pdiff{a}{t}\bigg|^2  + \frac{1}{2} \left( k^4+ 1\right)|a|^2  \D{k}\right]
     =  - 2\mu  \int^{\infty}_{-\infty}  |k| \left|\pdiff{a}{t}\right|^2 \D{k},
\end{equation}
where the term involving $F$ has vanished by virtue of the integral constraint (\ref{eq:a_k_integralconstraint}). 

Equation (\ref{eq:EnergyBudgetv3}) is an energy budget\footnote{It is noted that it is possible to arrive at the energy budget (\ref{eq:EnergyBudgetv3}) by means of an alternative \textit{physical approach}, in which one can derive an energy equation for $w$ before substituting in the Fourier decompositions (\ref{eq:FT_phi})--(\ref{eq:FT_w}) from the main body.} for the system (\ref{eq:ak_equation})--(\ref{eq:a_k_integralconstraint}). The left-hand side of (\ref{eq:EnergyBudgetv3}) represents the change in total energy within the system. From left to right the terms in the integrand represent the modal contributions to: the kinetic energy (density times the square of velocity); the bending cost of the beam in response to the fluid; and the substrate cost (cost of moving the fluid surface against pressure). 

The absence of the compressive force in (\ref{eq:EnergyBudgetv3}) is a reflection that --- by inextensibility --- there is no energetic penalty associated with a change in beam length. The right-hand side of (\ref{eq:EnergyBudgetv3}) represents losses in energy due to viscous dissipation. We refer to the terms within the integrand of (\ref{eq:EnergyBudgetv3}), which are given by
\begin{equation}\label{eq:AppendixEnergyExpressions}
    \mathcal{E}_{\mathrm{kin}} (k,t)= \frac{1}{2|k|} \bigg|\pdiff{a}{t}\bigg|^2
    \qquad
    \text{and}
    \qquad
    \mathcal{E}_{\mathrm{pot}} (k,t) =\frac{1}{2} \left( k^4  +1 \right)   |a|^2,
\end{equation}
as the kinetic and potential energy densities (energy per unit wavenumber). The sum of these energy components defines the total energy density
\begin{equation}
     \mathscr{E}(k,t) =  \mathscr{E}_{\mathrm{kin}}(k,t) + \mathscr{E}_{\mathrm{pot}}(k,t).
\end{equation}
The system's total kinetic and potential energy may then be obtained by integrating over all wavenumbers: 
\begin{align}
   E_{\mathrm{kin}}(t) = \int^{\infty}_{-\infty} \mathcal{E}_{\mathrm{kin}} \D{k} ;
   &&
   E_{\mathrm{pot}}(t) = \int^{\infty}_{-\infty} \mathcal{E}_{\mathrm{pot}} \D{k} ;
   &&
   E(t) = \int^{\infty}_{-\infty} \mathcal{E} \D{k} .
\end{align}
The energy budget \eqref{eq:EnergyBudgetv3} can now be expressed more succinctly as
\begin{equation}\label{eq:EnergyBudgetv4}
    \diff{E}{t} =  - 2\mu  \int^{\infty}_{-\infty}  |k| \left|\pdiff{a}{t}\right|^2 \D{k} .
\end{equation}
Hence, when $\mu=0$ the energy of the beam is conserved, as expected, otherwise it is non-increasing with non-zero velocities leading to energy losses through viscous dissipation.
\color{black}

\end{document}